\documentclass[9pt]{article}
\usepackage{graphicx}
\usepackage{epstopdf}
\usepackage{dcolumn}
\usepackage{bm}

\title{ Spooky black holes : gravitational self energy  and space curvature}

\begin{document}

\author {Paolo Christillin \\
Dipartimento di Fisica, \\
Universit\`a di Pisa}

\maketitle

\begin {abstract}  

It is shown that space curvature can be disposed of by   properly taking into account gravitational self energies.  This leads to a parameter free modification of Newton's law, violating Gauss theorem, which accounts for the crucial tests of gravitation in a flat space. Strong gravitational fields entail  opposing big gravitational self energies.  The negative gravitational self energy of a gravitational composite object, which results in a mass defect with respect to the sum of the constituents, thus cancels out the latter at the Schwarzschild radius. Hence a black hole, possible end result of the radiative shrinkage of a star, having zero total energy cannot any longer interact with other objects. Baryon number non conservation may result  .

\end {abstract}  
\

PACS numbers : 04.80Cc , 04.50.Kd , 04.60. -m , 04.70.Bw , 98.80.-k , 98.80.Bp

\

\section{Black or no holes ?}\label{}

\

There is no need to further underline the physical and cosmological relevance of black holes. Some speculations about them may hence be of some relevance.
Let us start with the usual elementary considerations using a classical (Newtonian) gravitational language. The escape velocity of a body of mass m, at the surface of another (big) celestial body of mass M and radius R, to reach infinity where the potential energy has been put to zero, is

\begin{equation}\label{R_{S}}
 1/2 \; m v^2 - GMm/R= 0
 \end{equation}

Following Laplace and Michell [1] , when however the mass M shrinks to dimensions of the Schwarzschild radius 

\begin{equation}\label{R_{S}}
 r_{S} = 2 G M /c^2
\end{equation}

which corresponds to an escape velocity v = c , not even light can escape the gravitational field. This defines the radius of a black hole, and , as well known, it would correspond to extraordinarily abnormal densities ; for instance the earth radius should reduce to about a millimeter ($R_{T} / r_{S} \simeq 10^{-9}$).

The previous argument is questionable in  two respects : first it uses a non relativistic expression for the particle kinetic energy, second it uses the non relativistic (Newtonian) expression for the gravitational interaction. 

However Eq.2) is said to be correct because  General Relativity  (GR) (\cite{Einstein}) would compensate the two mistakes . Indeed in the Schwarzschild metrics the time interval $ dt^2 $ gets multiplied by the factor $ (1 - r_{S}/r  )$ which does the job, 
predicting a red shift for a photon angular frequency $\omega'$ at a distance r'  from M, with respect to its value $\omega$ at the surface

\begin{equation}\label{R_{S}}
 \hbar \omega \sqrt {1   -  2GM/ c^2 R}  = \hbar \omega' \sqrt {1   -  2 GM/ c^2 r'}  
 \end{equation}

 Naive energy conservation would simply require 

\begin{equation}\label{R_{S}}
  \hbar \omega (1   - GM/ c^2 R) = \hbar \omega' (1   - GM/ c^2 r')  
  \end{equation}

It is immediate to see that no emission is possible when $ GM/ c^2 R = 1 $ i.e. at  $  {r'_{S} }= G M /c^2$ ,
a halved Schwarzschild radius. Notice that this is a straightforward extension (justified later) of the same argument, based on the mass-energy equivalence for the photon, used by Einstein himself at the surface of the earth (i.e. for the weak field case)
and that the two expressions agree to the first order . 

The common ingredient in both approaches is the Newtonian potential, which is then extrapolated to extreme conditions. Indeed the Schwarzschild solution  is matched to the external Newtonian solution in the weak field case  to connect the factor 2 to 1.

Consider now  the same problem for a mass m, treated relativistically  : when cannot escape and reach $\infty$ ?  Energy conservation  reads 

\begin{equation}\label{true}
 m_{0} / \sqrt {1 - v^2/c^2} \: (1   - GM/ c^2 R) = m_{0}
 \end{equation}

and again  also a particle of energy $ E = m_{0} c^2  / \sqrt {1 - v^{ 2}/c^2}   $    cannot escape a star at $  r'_{S}= r_{S}/2$ .

  Notice that only the fact of having used the relativistic expression for the energy (and the corresponding mass in the Newtonian potential)  has allowed the factorization. 
  
Let us then come to the physical reason for that and to an alternative interpretation of why nothing can escape a black hole.

We are going to show that strong field implies big opposing gravitational self energy.

We hence start with the simpler case of two gravitational bodies  M and m in the ordinary weak field situation, 
assuming for simplicity  that $M >> m $ (which represents e.g. the case for the solar system).

\section{Self energy vs. space curvature}\label{}

\

The gravitational interaction energy is traditionally given by the Newtonian expression

\begin{equation}\label{R_{S}}
E_{G} = U = -  G M m /r                          
 \end{equation}
  
  However, since {\it {\bf energy} is the source of gravitation} , the mass (energy)  m is not the mass the second object possesses at 
  $\infty$.  
  The interaction renormalizes it in a space dependent way 
  
\begin{equation}\label{R_{S}}
m => m' = m_{0} (1 -  G M  /c^2 r  )                        
\end{equation}

where $m_{0} $  stands for the bare mass (without the influence of M) , so that, relying again on Einstein's mass-energy equivalence (and disregarding special relativity effects) ,

\begin{equation}\label{R_{S}}
U' = -  (G M  m_{0}/ r)(1 -  G M  /c^2 r  )                           
\end{equation}
  
  or 
  
  \begin{equation}\label{phi'}
\phi' = -  (G M /r)/  (1 -  r'_{S}/r  )                           
\end{equation}

\

This is, mutatis mutandis, the gravitational analogue of mass renormalization in the e.m. case, where virtual processes have a different effect for a free and a bound electron, with an ensuing $\it {measurable} $ differences (of two  $\infty$s).

Here, apart from (non) quantization, the gravitational interaction has decreased the total energy of the system and hence of its mass which we approximately attribute to the lighter one (but remember that also in the Schwarzschild solution space is supposed to be curved only by the heavier mass. In the case of binary systems the reduced mass should probably intervene.) , modifying in turn the interaction.

This represents a low brow implementation of the {\it  non linearity of gravitation }.

It is then clear that the round bracket in Eq.s (4) and (5) incorporating energy conservation, should  be modified and go into 

  \begin{equation}\label{R_{S}}
1 -GM/c^2 r (1-GM/c^2 r) = 1 -x (1-x)                        
\end{equation}

The modification due to the self energy term can be interpreted in two alternative ways.

First, one realizes that Eq.(\ref{phi'}) can be recast into a modified  "effective" potential with an additional small exponent (e.g. for the Sun at Mercury $  G M  /c^2 r  \simeq 10^{-8} $ )

  \begin{equation}\label{R_{S}}
\phi' = -  G M /r^{(1+\alpha)}                          
\end{equation}

  where the {\it repulsive} extra term modifies  {\it in a parameter free way } the $1/r^2$ power law of Newton's law, which then reads
  
  \begin{equation}\label{R_{S}}
{\bf  F} = -  G M m \;  {\bf r}  /r^3 \; \: (1- 2 r'_{S}/r)                        
\end{equation}
  
   The role of the extra term becomes more relevant at short distances, which makes it plausible why only in the case of Mercury it has played some role. 

{ \it In turn this implies a violation of Gauss's theorem, if  expressed as usual in terms of a given 
 constant mass.}

But, probably more interesting, one can also define 

  \begin{equation}\label{R_{S}}
\phi' = -  G M /r'                        
\end{equation}

where

  \begin{equation}\label{R_{S}}
r'  = r /  (1 -  G M  /c^2 r  )  \simeq r/ \sqrt{   (1 -  r_{S}/ r  )  }                      
\end{equation}

where  the last provocative passage holds true up to second order terms in $G M  /c^2 r $.

Notice however that  no singularity appears  since the renormalized mass can be zero at worst in the defining equations. 

The correction factor $x = 1 -  G M  /c^2 r  $ intervenes in a two-fold way. 

 In a first instance one keeps the usual Newtonian potential for  the standard gravitational attraction. Just because of energy conservation one predicts the first order gravitational red shift . Then self energy effects are taken into account. They can be disposed of in the traditional language just by rescaling, as above mentioned, $dt$  and $1 / dr $ , by the factor $ 1- Gm/c^2 r $, hence causing a " first order   curvature of space- time ".   
  
 But one can also rescale everything by $ 1-x (1-x)$ .  This would correspond to have eliminated also the interaction, as in Einstein's approach , expressed by
the Schwarzschild invariant interval 
 
 \begin{equation}\label{R_{S}}
(ds)^2 = (1-r_{S}/r ) (cdt)^2 -  (dr)^2 / (1-r_{S}/r) -r^2 (sin^2 \theta (d\phi)^2 + (d\theta)^2) 
\end{equation}

or in other words : { \it space curvature is due to neglect of self gravitational effects }.

Note parenthetically that also the previous expression, as well known ,  is not singular at  $r_{S}$  since \cite {Fey},  by the change of coordinates  $r = R (1+r_{S}/4R)^2 $ ,   can be transformed into
 
 \begin{equation}\label{R_{S}}
(ds)^2 = (1-r_{S}/4R)^2 /(1+r_{S}/4R )^2 (cdt)^2 - (1+r_{S}/4R)^4 ((dx)^2 + (dy)^2 + (dz)^2) 
\end{equation}

  where $R^2 = x^2 + y^2 + z^2$  .   As regards the time coefficient,  $R= r_{S}/4$ corresponds to  $r = r_{S}$ so that the two expressions cancel out at the same value and agree at $\infty$  as well as with the Newtonian limit.

  Finally , as it will be commented upon , second order terms are at variance with the Schwarzschild predictions.

Therefore, considering that  the traditional  tests of GR  can be explained in the Appendix without its full machinery , the previous arguments  probably deserve some consideration.

Let us then pass to consider composite objects and come back to black holes.

There is little doubt that the gravitational energy comes from the elementary interaction of the particle m with the individual masses $\it m_{j} $ which make up the mass $M = \Sigma_{j} m_{j} $ with the previous choice of the zero of the potential, with respect to the situation where all of them are at rest at infinity. 

It is worth stressing that this must hold true also for the mutual interaction of the masses $m_{j} $ .
Indeed the gravitational self energy of M in the case of constant matter density $ \rho = M / (4/3 \pi R^3)$ is given by 

\begin{equation}\label{R_{S}}
 E_{G} = - (3/5)   G M^2 /R                          
  \end{equation}

which, when compared to the relativistic energy $M c^2$ of the body M, yields 

\begin{equation}\label{R_{S}}
 E_{G}/Mc^2  = - (3/10)  r_{S}/R
\end{equation}

 totally negligible under normal conditions.

Manifestly this does no longer hold true under extreme conditions and it is paramount to determine exactly the previous ratio. 

In this connection let us first stress that the linear relation between $r'_{S}$  and  ${M'}_{S}$ cannot be accounted for by a constant         $ \rho(r) $.  Such an uncommon configuration can be realized with $ \rho_{sing}(r) = \rho _{o}/(4 \pi r^2)$.   In this case 

\begin{equation}\label{R_{S}}
 M'_{S} =  \int \rho(r)_{sing}  d^3 r = 4 \pi\int \rho (r)_{sing}\ r^2 dr =   \rho _{o}r'_{S} =(  c^2/G )  r'_{S} 
 \end{equation}

Therefore 
 
 \begin{equation}\label{R_{S}}
 E_{G} =  - \int  G  ( \rho_{o}  r )  ( \rho_{o} dr ) /r  = - G  \rho _{o}^2 r'_{S}  = - G  M^{' 2}_{S} / r'_{S} = - M^{' }_{S} c^2
 \end{equation}

 We thus see that  the  value at which neither light nor a mass m  could escape 
 
 \begin{equation}\label{R_{S}}
 (1   - GM/ c^2 R) = 0 
  \end{equation}  
 
  corresponds to value where the gravitational self energy has swallowed the mass M of the gravitational source which does no longer exist !

 \
 
\begin{figure} [h]
\centering
\scalebox{1.40}{\includegraphics{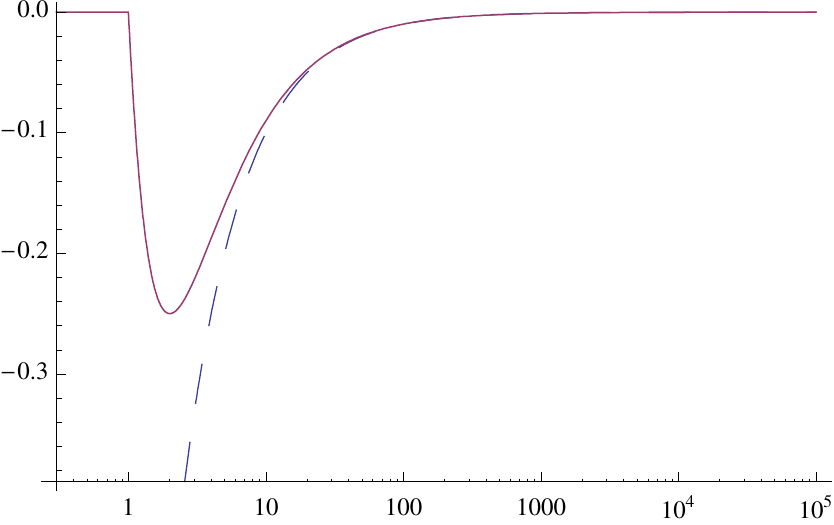}}
\caption{ Gravitational potential of an object of given mass M as a function of  $ r'_{S}/r  = GM / c^2 r $ . Dotted line usual Newtonian potential, solid with self energy. Deviations become sizable only at neutron star level.}
\label{black}
\end{figure}

 We can thus reconcile the two seemingly unrelated effects of renormalization of a mass m due to gravitation with a final cancellation at $r'_{S}$ and the disappearance of its gravitational source M at the same  $r'_{S}$ .   Indeed m can reach smaller and smaller distances only if M shrinks more and more. 
In the end the distance where the {\it total}  gravitational energy of m disappears so that 

 \begin{equation}\label{R_{S}}
 m_{\infty} = m_{0}( 1-x (1-x) ) => m_{0}
 \end{equation}

 and m feels nothing corresponds to the distance where the self energy has swallowed M   Fig.\ref{black} ) .
 We have therefore a twofold confirmation of the non existence of black holes.

\

\

  Let us now proceed to the same calculation within the formalism of GR.
  
  This time there will be a factor of 2 in the definition of $\rho_{0}$
but
 the  volume element, because of the non Euclidean space, gets modified in the Schwarzschild metrics  \cite {Wald} by $ 1/ \sqrt {1-r/r_{S}}$ .  

Thus the self energy contribution will be 
$2 M_{S} ^2 /r_{S}$ again equal to $M_{S} c^2$  , with a cancellation at the Schwarzschild radius (remember  the difference between primed and unprimed quantities).

 In the standard GR formalism this has been partially recognized   with the observation  (without saying anything definite about $\rho$ ) that  $E_{B} = M_{P}   - M $, where

 \begin{equation}\label{R_{S}}
 M_{P} =  \int \rho(r)  d^3 r / \sqrt {1-r/r_{S}} > M 
 \end{equation}

  might  " be interpreted as the gravitational binding energy of the configuration"  \cite {Wald} .  

 Therefore some confidence in our result might be reasonable since  the same cancellation mechanism 
 against the relativistic mass seems to operate  
, consistently with the no escape considerations,  in both approaches  (the difference between $r_{S}$ and $r'_{S}$ compensating for the curvature vs. flatness in the cancellation).  In addition the exact expression of the critical radius seems to be irrelevant to the argument. Within a factor of two around the critical value the gravitational  self energy would completely swallow the initial mass of the star. 

This is the  gravitational analogue of the chemical and nuclear mass defect. Binding decreases the "mass" of a structure with respect to the sum of its constituents; all the more so the stronger the binding. And a black hole would represent the astonishingly strongest gravitational example : the whole mass having been eaten up by binding. Apparently astonishingly because being provided by the weakest interaction  thanks to its long range feature and, unlike e.m. , because of its additivity , but long well known to the astrophysical community.
Collapsing AGN 's are indeed known to provide  large amount of radiation of the order of 20 per cent  of their mass : gravitation represents therefore the most efficient engine to furnish energy.   
 Indeed the rest mass of an object is an irrelevant quantity as long as the object does not decay. Here, in  a sense, gravitation would induce such a process.

Of course we cannot be completely sure of what happens at these extreme conditions although collapse seems to be reasonable \cite{Lan} \cite{foot} .

 As regards their actual presence (i.e. not just a mathematical solution) , to prove or disprove the present considerations, unless one has a separate reliable measure of both R and M, all talks will be just metaphysical. 
 
  In this connection the "presence" of super massive black holes of  mass $10^6 - 10^9$  solar masses at the center of galaxies (for ours the first figure would apply) should be questioned according to the present considerations.   As a matter of fact if they were true black holes they would no longer exert any attraction so that there would be no explanation for the stars orbits. To account for that,  a highly speculative proposal (not more unfounded than postulating an unknown mechanism for generating such massive doubly  exotic objects ) , in agreement with other formation mechanisms, might be a cluster of neutron stars.

In conclusion an orbiting mass m around a Schwarzschild shrinking star, could we observe its evolution,  would thus eventually "unexplainably  escape " , accompanied by strong x ray  and gravitational waves emission from the star which would finally disappear.

In other words the gravitational field of a collapsing gravitational object becomes weaker and weaker, eventually disappearing at the Schwarzschild radius.

\

 The energy has of course not disappeared but has been rather radiated away in the process, baryon number being in the end no longer conserved. 
 As can be seen from Fig. 2) it does not seem that the strong field limit be attained in nature.

\

\begin{figure}   [h]
\centering
\scalebox{0.40}{\includegraphics{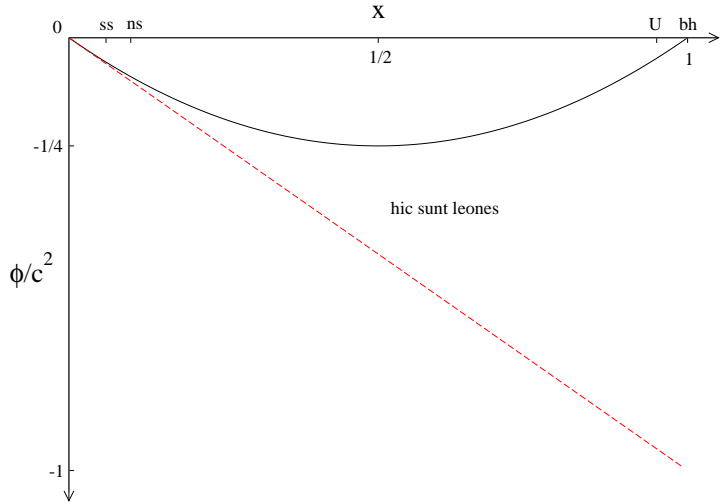}}
\caption{The gravitational potential $\phi /c^2 $ as a function of $x= GM/c^2r$ at the surface of different gravitational objects.
Broken line without  , continuous with self energy  i.e. $ \phi /c^2 = - x(1-x)$ . Symbols : ss stands for solar system, ns for neutron star,  bh for black hole and U for Universe.  $ \phi /c^2 \leq 1/4$ hence everywhere gravity should  remain essentially weak . Not in scale }
\end{figure}

The thermodynamical treatment of black holes as well as of their evaporation seems therefore superseded.

\section{Conclusions}\label{}

A  pedestrian revisitation of gravitation has been attempted. 

It has been  shown that strong a gravitational field entails at the same time an opposing strong self energy effect. 
This makes less dramatic (and plausibly null)  the effect of black holes on other bodies.
The same effect is of course operative in the two body case, where again physically unavoidably  self energy effects which renormalize the post Newtonian gravitational interaction provide an alternative scenario which disposes of space curvature  so that even in the presence of gravity one can work with a Minkowski invariant four-interval i.e. in a flat pseudo-Euclidean spacetime.  Thus the velocity of light remains constant even in the gravitational case. 

Of course the local equivalence of gravitational and inertial mass, which is the essentials of Einstein' legacy, remains unquestioned.

It thus appears worth considering  whether the extension  of  Newton's static expression, which has been used throughout with self energy corrections,  to

  \begin{equation}\label{R_{S}}
(1/c^2 \partial ^2/\partial t^2 - \nabla) \phi = 4 \pi  G \rho / c^2                       
\end{equation}

to account for finite time propagation \cite {Zeld} (although retardation effects are known to play an even lesser role that in e.m. {\cite {footend} ) ,  within a covariant formulation of {\bf F} = m {\bf a},  might provide an alternative calculative  more viable way to describe gravity .

\
\

\section{Appendix}\label{}

\

As recalled in the text, just because of the mass-energy equivalence, together with the extension to any distance of the on earth photon gravitational attraction (used by Einstein and confirmed by the Pound-Rebka \cite{pound} and Briatore-Leschiutta \cite{bria}  experiments)

  \begin{equation}\label{R_{S}}
 \Delta t' / \Delta t  = (1   - r'_{S}/ r)/1   - r'_{S}/2 r')  
\end{equation}

(here  r' just  refers to another different distance in the gravitational field).
We can thus think of 

  \begin{equation}\label{R_{S}}
 \Delta \tau_{G} =  \Delta t  /1   - r'_{S}/ r)  
\end{equation}

time measured at a distance r in the gravitational field of a body M, as the proper "gravitational" time, with respect to a hypothetical "absolute" time at $\infty $ , (we disregard academic speculations about surrounding hollow sphere which change the potential which do not seem anyhow to have any physical effect on the differences we measure),  the perfect analog of

\begin{equation}\label{R_{S}}
 \Delta \tau_{R}  =   \Delta t  / \sqrt{1   - v^2/c^2 } 
\end{equation}

of special relativity.
It has also been shown that the undue neglect of physically founded non linear self-energy effects can be cured by simply rescaling distances as 

  \begin{equation}\label{R_{S}}
 r'  =  r  /(1   - r'_{S}/ r)  
\end{equation}

In other words, the distance r we would measure with another non gravitational probe (e.g. an electric one, allowing M to be also charged) turns, because of the non linearity of gravitation, into r'. 

\

It is then immediate to account for the so called "crucial tests of GR".

\

a) Gravitational red shift.

\

Trivial consequence of gravitational energy conservation.  As stressed in the text  differences among the different formulations would entail a difference in the predictions  for GPS observations of $\simeq 1 \:  cm$ in one year. Indeed the present formalism predicts as a second order effect $ + x^2$ as compared to $ - x^2/2$ of the Schwarzschild solution.
This $ O(10^{-19})$ effect, if measurable, would really constitute a stringent test (definitively more than the $ O(10^{-8})$ precession of Mercury's perihelion)  of the correct approach. 

\

b) Light deflection.

\

As in the following cases, the routine ingredients to solve these two body planar problems are angular momentum and energy (already commented upon) conservation. 

We consider as usual a  luminous  ray grazing the sun, coming from $\infty $ and calculate the  {\bf light  deflection} at $R_{S}$. The deflection measured by a distant observer (we on the earth, which is also practically at $\infty $,  so that its location does not intervene ) will be just twice  as much.

The angular momentum (which we  denote by the traditional L) must be obviously conserved.  The quantities entering L at the  Sun  are changed as seen from the earth because of the previous relations.  Hence  the photon which was locally assumed to be perpendicular, is necessarily perceived to deviate by an angle $ \Delta  \theta'$

\begin{equation}\label{psi}
c L /\hbar =  \omega_{S}   R_{S}  \rightarrow  (1   + GM/ (c^2 R_{S}))  \omega_{T} R_{T}  / (1   - GM/ (c^2 R_{S}) )    \Delta  \theta'
\simeq ( (1   + 2 GM/ c^2 R_{S}) \omega_{T} R_{T}  \Delta  \theta'
 \end{equation}

 where the small angle approximation has been made . 
 
 Thus  the final  deflection is given by  by  $ \Delta  \phi'  = 2 \Delta  \theta' $ or 
 
 \begin{equation}\label{psi}
\Delta \phi'  =  - 4 G M_{S} /(c^2 R_{S}) 
 \end{equation}

the minus sign meaning that the photon must be of course attracted by the Sun.

Notice that there is no light velocity dependence on gravity  ($ c' = c_{0}( 1+ 2 \phi/c^2) $)  in the present  approach. The photon of energy $\hbar \omega$ is attracted as any other energetic (massive) object and its momentum enters angular momentum conservation as $ p = E/c = \hbar \omega/c$  (in other words physically $\hbar \omega'/c  \neq \hbar \omega/c' $). Hence the photon keeps  propagating in {\it any}  gravitational field with speed c, once the non linearity of gravitation has been correctly taken into account.

\

c) Radar time delay

\

The undisturbed straight line light trajectory is deflected by the presence of the Sun (we defer e.g. to \cite {Gasp} for the symbols). Along this trajectory time and space are affected by gravity as before, resulting in a longer trajectory along which time runs slower.   Hence for us the signal suffers a retardation given by

 \begin{equation}\label{psi}
dt   =   2 G M_{S} /c^2 r  ( dx/c )
 \end{equation}

x standing for the unperturbed trajectory coordinate (of total length $x_{P} - (-x_{T}) $ , at a distance R from the Sun )  between the planet  and the earth  at radial distances distances  $R_{P} , R_{T} $ from the Sun ) and r for the actual one. 
Integration of the previous expression yields the known result for the delay

  \begin{equation}\label{psi}
\Delta t   = 4 G M /c^3  \times ln ((R_{P} + x_{P})/(R_{T} - x_{T}))
 \end{equation}

As stressed before the speed of light c is constant  .

\

d) Precession of Mercury's perihelion.

\

This time again proper angular momentum conservation yields the extra "torque" which makes the ellipse to rotate and advance. 
The standard expression  

 \begin{equation}\label{psi}
d\phi/dt   =  L/r^2
 \end{equation}

 for the above mentioned reasons  gets an extra $r'_{S}/  r $ factor from the gravitational time correction and   
two from the square of the radius .  This results in the appearance  in the radial equation 

\begin{equation}\label{psi}
d^2 r /d\tau_{R} ^2  = -GM/(c^2 r^2) + L^2/r^2 (1-3 GM/c^2r))  
 \end{equation}
 
 and the standard expression for the  precession

\begin{equation}\label{psi}
\Delta  \omega' /\omega '  =  \Delta  \phi' /\phi'   = 6 \; G M_{S} /(c^2 a (1-e^2) ) 
 \end{equation}

 follows along standard lines.
Q.E.D. .

In terms of the factor  $GM/(c^2 r)$ the coefficients 2 and 3 entering respectively light bending and the perihelion precession have  a transparent meaning. 

{\it It is not superfluous  to underline why the despised proposal of modifying Newton's law in an ad hoc manner was also successful for Mercury :  the present parameter free treatment  justifies the equivalence of the two approaches ! }

\

The order of magnitude of the effects considered  is thus : 

light deflection by the sun  $ 2 GM_{S}/c^2 R_{S}  \simeq 10^{-6}$ 

 precession of the perihelion of Mercury $  6 G M_{S} /(c^2 a (1-e^2) ) \simeq 10^{-8}$

 gravitational violet shift for GPS $\: $  $  GM_{T}/c^2 R_{T}  \simeq 10^{-10}$, which can be further lowered to $ O(10^{-15}) $ by the factor $h/ R_{T}$ in the Pound-Rebka experiment and to $ O(10^{-19}) $ for yearly observations. 

It is then self evident that the most stringent  tests of gravitation take place, by far, on the earth , thus  lessening the importance of  Mercury precession. 

\

\

ACKNOWLEDGMENTS

\

It is a pleasure to thank P.G. Prada Moroni for his patience in listening to me, for constructive criticism and for fruitful comments .
I also wish to thank  G. Morchio ,   M. Velli and   M. Lucchesi     for helpful discussions and for a critical reading of the manuscript.

\ \\


\begin{thebibliography}{999}
\bibitem{foot}
One can probably gain some additional confidence by considering an electromagnetic analogue, that of positronium, which is also a system with negative heat capacity. The electron positron pair of initial mass 2m starts spiraling, RADIATING  so that its  energy is given by $2m c^2 -e^2 /r $.  In this process the velocity of the pair increases as well as its negative self-energy. Finally   at  $r = e^2/(2m c^2) $, which is the classical radius of the electron,  it annihilates (here the factor 2 coming from the reduced mass is irrelevant to the argument as in the gravitational case) . The classical radius represents for the composite electrical system what the Schwarzschild radius represents, mutatis mutandis, for a composite gravitational object.   Energy has been conserved although of course not locally ; the { \it gravitational } energy of the pair which was given at the beginning by the two masses will be zero in the end having been eaten up by the self energy .
\end{thebibliography}
\end{document}